\documentclass[apjl]{emulateapj}         
\usepackage{apjfonts,graphics}

\newcommand{\um}{${\rm \mu m}$~}
\newcommand{\mm}{${\rm \mu m}$}

\long\def\symbolfootnote[#1]#2{\begingroup%
\def\thefootnote{\fnsymbol{footnote}}\footnote[#1]{#2}\endgroup}

\slugcomment{accepted for publication in \apj} 

\begin{document}

\title{ASTEROID BELTS IN DEBRIS DISK TWINS: VEGA AND FOMALHAUT}
 \author{Kate Y. L. Su\altaffilmark{1}, 
   George H. Rieke \altaffilmark{1},
   Renu Malhotra\altaffilmark{2},
   Karl R. Stapelfeldt\altaffilmark{3},
   A. Meredith Hughes\altaffilmark{4,10},
   Amy Bonsor\altaffilmark{5},
   David J. Wilner\altaffilmark{6},
   Zoltan Balog\altaffilmark{7},
   Dan M. Watson\altaffilmark{8},
   Michael W. Werner \altaffilmark{9},
   Karl A. Misselt\altaffilmark{1}
  }

\altaffiltext{1}{Steward Observatory, University of Arizona, 933 N Cherry Avenue, Tucson, AZ 85721, USA; ksu@as.arizona.edu}
\altaffiltext{2}{Lunar and Planetary Laboratory, University of Arizona, Tucson, AZ 85721, USA}
\altaffiltext{3}{Code 667, NASA Goddard Space Flight Center, Greenbelt, MD 20771, USA}
\altaffiltext{4}{Department of Astronomy, University of California, Berkeley, CA 94720, USA}
\altaffiltext{5}{UJF-Grenoble 1/CNRS-INSU, Institut de Plan\'etologie et d'Astrophysique de Grenoble (IPAG), UMR, 5274, Grenoble, F-38041, France}
\altaffiltext{6}{Harvard-Smithsonian Center for Astrophysics, 60  Garden Street, Cambridge, MA 02138, USA}
\altaffiltext{7}{Max-Planck-Institut f\"ur Astronomie, K\"onigstuhl 17 D-69117, Heidelberg, Germany}
\altaffiltext{8}{Department of Physics and Astronomy, University of Rochester, Rochester, NY 14627, USA} 
\altaffiltext{9}{JPL/Caltech, 4800 Oak Grove Drive, Pasadena, CA 91109, USA}
\altaffiltext{10}{Miller Fellow}

\begin{abstract}

Vega and Fomalhaut, are similar in terms of mass, ages, and global
debris disk properties; therefore, they are often referred as ``debris
disk twins''. We present {\it Spitzer} 10--35 \um spectroscopic data
centered at both stars, and identify warm, unresolved excess emission
in the close vicinity of Vega for the first time. The properties of
the warm excess in Vega are further characterized with ancillary
photometry in the mid-infrared and resolved images in the far-infrared
and submillimeter wavelengths. The Vega warm excess shares many
similar properties with the one found around Fomalhaut. The emission
shortward of $\sim$30 \um from both warm components is well described
as a blackbody emission of $\sim$170 K. Interestingly, two other
systems, $\epsilon$ Eri and HR 8799, also show such an unresolved warm
dust using the same approach. These warm components may be
analogous to the solar system's zodiacal dust cloud, but of far
greater mass (fractional luminosity of $\sim$$10^{-5}-10^{-6}$
compared to $10^{-8}-10^{-7}$). The dust temperature and tentative
detections in the submillimeter suggest the warm excess arises from
dust associated with a planetesimal ring located near the water-frost
line and presumably created by processes occurring at similar
locations in other debris systems as well. We also review the
properties of the 2 \um hot excess around Vega and Fomalhaut, showing
that the dust responsible for the hot excess is not spatially
associated with the dust we detected in the warm belt. We suggest it
may arise from hot nano grains trapped in the magnetic field of the
star. Finally, the separation between the warm and cold belt is rather
large with an orbital ratio $\gtrsim$10 in all four systems. In light
of the current upper limits on the masses of planetary objects and
the large gap, we discuss the possible implications for their
underlying planetary architecture, and suggest that multiple, low-mass
planets likely reside between the two belts in Vega and Fomalhaut.

\end{abstract} 


\section{Introduction}

Debris disks were discovered by {\it IRAS} as infrared excess emission
arising from systems of particles analogous to enhanced Kuiper
belts. They are tenuous dusty disks sustained by cometary activity and
planetesimal collisions, which initiate cascades of further collisions
that break bodies down into fine dust \citep{dominik03}. The dust is
lost relatively quickly through photon pressure, Poynting-Robertson
(P-R) or stellar-wind drag forces, generally in a time scale much
shorter than 10$^4$ yr. Thus, a debris-disk-generated infrared excess
requires the presence of colliding planetesimals, and a larger object
(can be as small as Pluto-size) to stir them, i.e., some form of
planetary system \citep{wyatt08}.  The large emitting area of debris
makes these disks detectable through infrared/sub-millimeter thermal
emission or optical scattered light, providing insights into the
nature of unseen parent-body populations and massive perturbers around
other stars.

The dust around a host star absorbs stellar radiation and re-emits in
the infrared at equilibrium temperatures, balanced between absorption
and emission. Consequently, the spectral energy distribution (SED) of
the infrared excesses and its relation to the dust temperatures can be
used to infer the number of emitting grains and their distances from
the heating star. For example, excesses in the near-infrared near 2
\um should be dominated by dust at $\sim$1500 K, excesses in the
mid-infrared near 24 \um should be dominated by dust at $\sim$120--150
K, while excesses in the far-infrared are mostly from dust at $\sim$50
K. Furthermore, excess emission at shorter wavelengths also contributes
excess at longer wavelengths as Rayleigh-Jeans (a steep function of
wavelengths).

Identifying excess emission around a star requires a good knowledge of
stellar photospheric properties and extrapolation to long
wavelengths. Positive identifications of excesses at long wavelengths
are easier than at shorter wavelengths where the host star dominates,
unless the signal of the star can be filtered out through
interferometric techniques. There is also a concern whether excesses
at mid-infrared and far-infrared are tracing separate components since
the majority of debris disks are unresolved. For nearby systems,
modern facilities like {\it Spitzer} and {\it Herschel} have
sufficient resolution to resolve the detailed structures of the disk
and reveal the complexity in disk structures in a few cases. The
identification of a warm excess in these resolved systems requires
precise subtraction of the stellar photosphere in the resolved images,
which has been done for the Fomalhaut \citep{stapelfeldt04},
$\epsilon$ Eri \citep{backman09}, and HR 8799 \citep{su09}
systems. Ironically, some of the nearby resolved systems are saturated
in the mid-infrared, making the recognition of such a component very
challenging.

In this paper, we present mid-infrared spectroscopic studies centered
at the two nearby A-type stars Vega and Fomalhaut obtained with the
{\it Spitzer} Infrared Spectrograph (IRS) instrument and we identify
warm excess emission in the vicinity of the star, clearly separated
from the much brighter cold planetesimal belts that dominate the
far-infrared emission. We suggest the presence of a planetesimal belt
near the water-frost line of the Vega system for the first time. We
compare its properties with the warm excess around Fomalhaut
\citep{stapelfeldt04}. We discuss the properties of the warm excesses
complemented with resolved infrared and submillimeter images of both
systems. Two other spatially resolved debris systems, $\epsilon$ Eri
\citep{backman09} and HR 8799 \citep{su09}, are also known to possess
a similar warm belt that is fainter and clearly separated from the
more prominent cold belt using a similar approach.  These warm
components may be analogous to our asteroid belt, but of far greater
mass. We discuss the implications and origins of this two-belt
configuration in light of the similarity in dust temperature
distribution found around unresolved debris systems between solar-like
and early-type stars by \citet{morales11}.

The paper is organized as follows. The observations and data reduction
are described in Section 2, including both {\it Spitzer} IRS
spectroscopy and {\it Herschel} PACS imaging. Photospheric
determination using ancillary data is given in Section 3.1 where we
conclude that no significant excess is found from 2.2 to 8 \um for
both systems. Using the PACS 70 and 160 \um images, we estimate the
excess flux of the unresolved source centered at the star position for
Vega in Section 3.2 and Section 3.3 for Fomalhaut. We construct the SEDs of
the unresolved excess components along with other mid-infrared and
submillimeter measurements, and suggest that they arise from a
planetesimal belt placed near the water-frost line in Section 4. In Section 5,
we discuss the implication of the two-belt systems on the underlying
planetary configuration, and provide a possible explanation for the 2
\um excess in the Vega system. Conclusions are given in Section 6.

\section{Observations and Data Reduction} 
\label{obs}

\subsection{{\it Spitzer} IRS Spectroscopy}
\label{irsobs}

To avoid saturation, we only used
data taken in the IRS high-resolution (R$\sim$600) modules (short-high
(SH): 9.9--19.6 \um and long-high (LH): 18.7--37.2 \mm).  The sizes of
the slits are 4\farcs7$\times$11\farcs3 and
11\farcs1$\times$22\farcs3 for the SH and LH modules,
respectively; a significant fraction of the light from a point source
is outside the slit. Standard slit loss correction for a point source
was applied to the extracted spectra; therefore, the part of the
spectrum where a point source dominates the emission has the correct
spectral shape.

\begin{figure} 
  \figurenum{1}
  \label{irs_vega} 
  \epsscale{1.0}
  \plotone{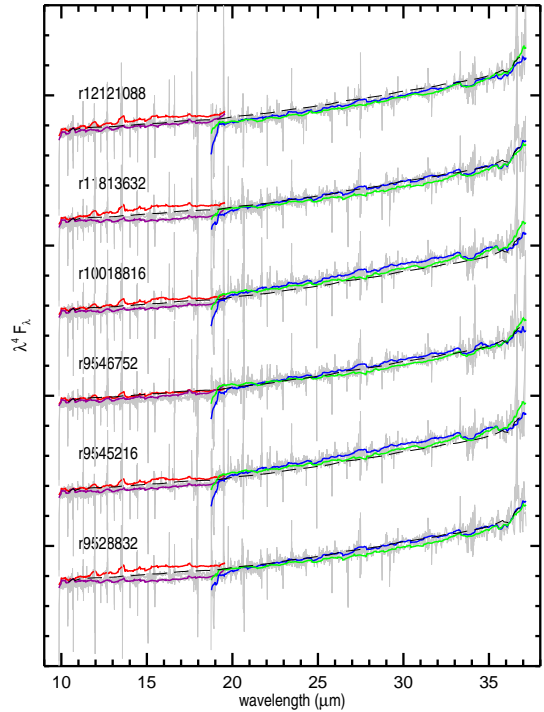}
  \caption{IRS spectra of Vega shown in $\lambda$ vs.~$\lambda^4
F_{\lambda}$ format so that a Rayleigh-Jeans spectrum is flat. Each observation
(AOR) is shifted vertically for clear viewing. The unsmoothed,
unscaled extracted spectra are shown in gray lines, and the scaled,
smoothed spectra are shown in colors: red and purple for SH while
blue and green for LH. The final combined spectrum (black dashed line)
is also shown in all six AORs for comparison.}
  \epsscale{1.0}
\end{figure}

IRS SH and LH spectral mapping data centered at the position of Vega
have been obtained through IRS calibration programs (PID 1406, 1409,
1411, and 1413) in 2004. Here we present six sets of observations where
the slit was placed on the Vega position based on PCRS pointing
information (no IRS peakup). These data were taken in the spectral
mapping mode with two positions parallel to the slit (7\farcs47 step$^{-1}$)
and three positions perpendicular to the slit direction (1\arcsec step$^{-1}$);
we only used the ones taken at the center perpendicular position (two
positions along the slit center). We used the SMART software (v.8.2.5;
\citealt{higdon04}) to reduce the BCD products from the {\it Spitzer}
Science Center (SSC) pipeline version of S18.18. Each of the spectra
was extracted with the full-slit mode without sky subtraction, flux
calibrated to a point source using the S18.18 calibration, and shown
in Figure \ref{irs_vega}. We then pinned all the SH spectra to 35.03
Jy at 10.6 \um according to the absolute calibration scale defined in
\citet{rieke08}, and shifted the corresponding LH spectra (a scale
factor of 0.975) to join smoothly with the SH ones. The final spectrum
was obtained as a weighted average of these six AORs (black dashed
line in Figure \ref{irs_vega}).

\begin{figure*} 
 \figurenum{2}
 \label{irs_vega_fom}
\plottwo{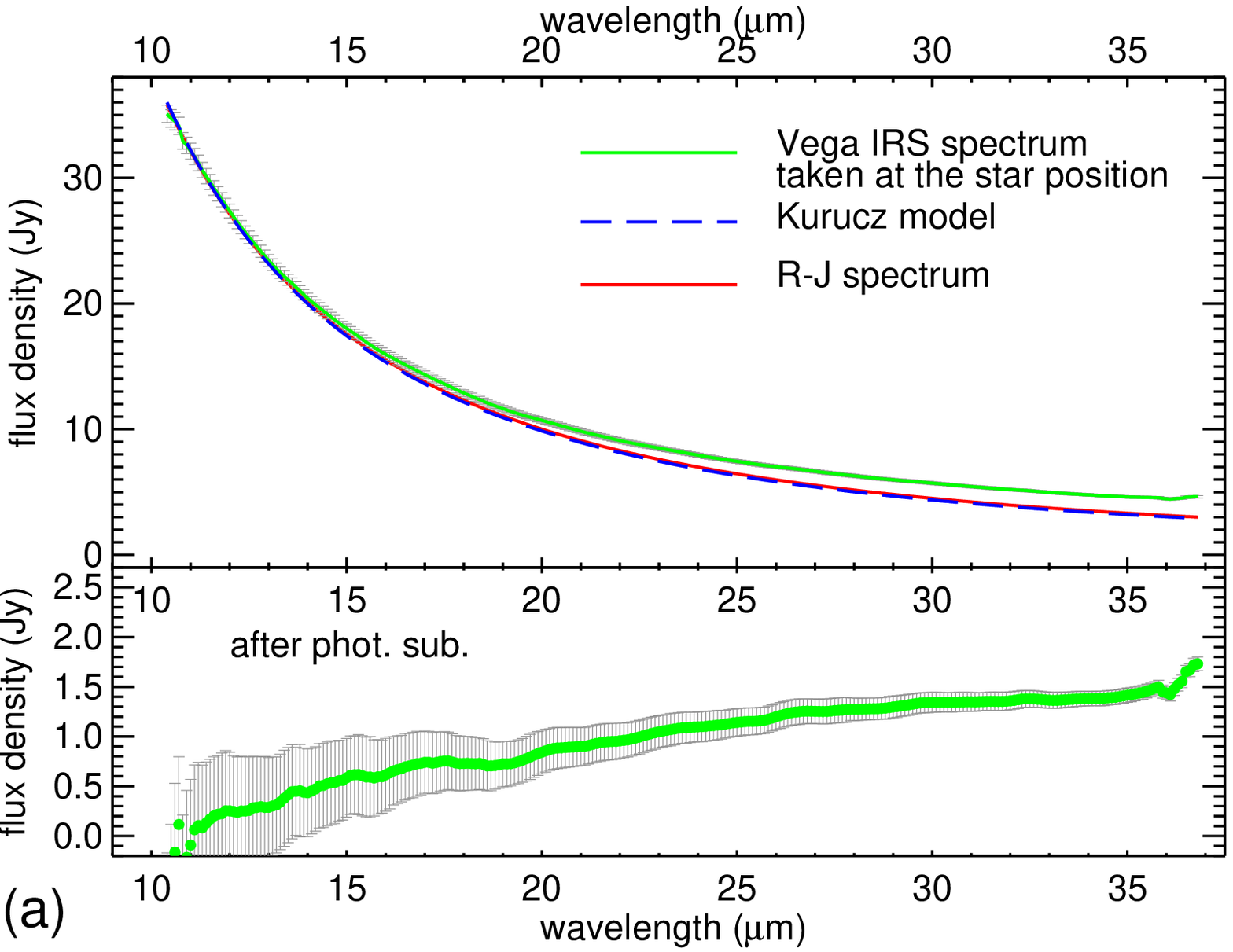}{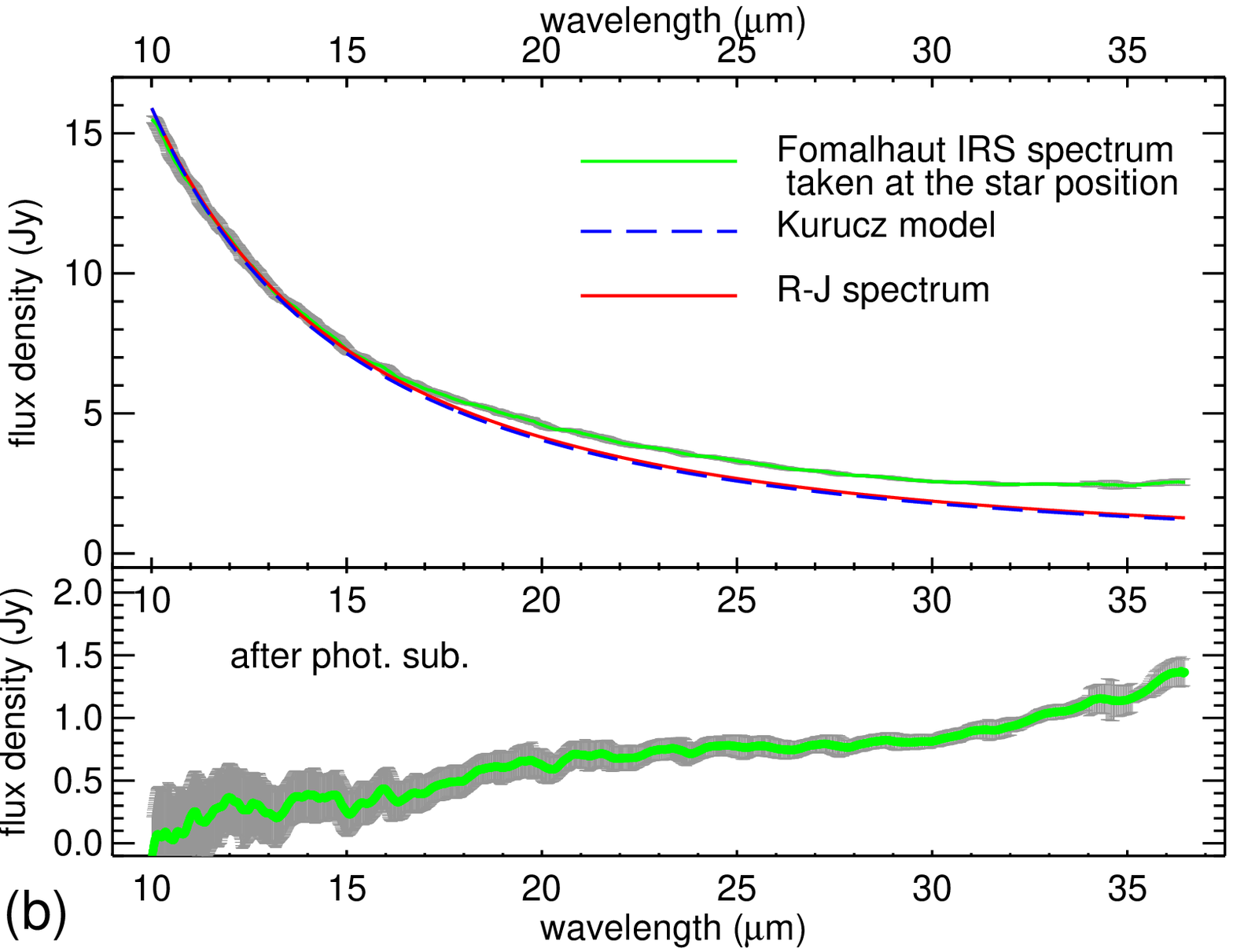}

 \caption{{\it Spitzer} IRS spectra centered at the star position for
Vega (upper panel a) and Fomalhaut (upper panel b). For comparison,
Kurucz atmospheric models (long dashed lines) and Rayleigh-Jeans
spectra (sold red lines) are also shown. The excess spectra (after
photospheric subtraction) are shown in the lower panels of the
plots. The errors for these excess spectra include 2\% uncertainty
from the photospheric models.}
 \epsscale{1.0}
\end{figure*}

The Fomalhaut system was observed with the {\it Spitzer} IRS
instrument in program PID 90 in 2004 June and November. Again, we only
report here the observations with the IRS SH and LH modules, due to
saturation of the signal in the low-resolution modules. In IRS SH, we
obtained a standard staring-mode nodded observation, with on-target
integration time of 503 s. This observation was preceded by a
high-accuracy pointing peakup on a nearby star with no infrared
excess, HD 216922, using the IRS Blue (13--19 \mm) peakup camera. With
IRS LH we obtained spectral-mapping observations with a strip of nine
slit positions separated in the dispersion direction by 4\farcs8\ and 
centered on Fomalhaut. The integration time per slit position was
122 s. The spectral mapping exercise was preceded by a
moderate-accuracy IRS pointing peakup using the red (19--26 \mm)
camera and HD 216922. Spectra at the extremes of the spectral map
indicated that sky emission is negligible compared to that by
Fomalhaut, so no sky subtraction was performed on the center-position
spectrum we discuss here. Data reduction began with basic calibrated
data products of the IRS data pipeline, version S11.  From these data
we removed, by interpolation in the spectral direction, permanently
bad and ``rogue'' pixels identified in the IRS dark-current data for
all observing campaigns up to and including the one in which each
Fomalhaut spectrum was taken. Again we used SMART for full-slit
extractions of spectra from the two-dimensional data. Similar
observations were made of $\alpha$ Lac (A1 V) in IRS SH and $\gamma$
Dra (K1 III). Along with template spectra for these stars provided by
M. Cohen (2004, private communication), we used these observations to
construct relative spectral response functions (RSRFs) for each
spectrometer, and in turn used these RSRFs to calibrate the Fomalhaut
spectra. We estimate the resulting spectrophotometric accuracy to be
approximately 5\%.

The final combined, smoothed spectra for both systems are shown in
Figure \ref{irs_vega_fom}. For comparison, the stellar photospheric
models (details see Section \ref{stellar}) are also shown. Given the
geometry of the outer cold rings (radius of 20\arcsec~in the Fomalhaut
system \citep{acke12} and radius of 11\arcsec~in the Vega system
\citep{sibthorpe10}) and the sizes of Point Spread Functions (PSFs) 
in the IRS wavelengths, the
spectral flux beyond $\sim$30 \um is partially contaminated by the
cold ring. The spectral shape and flux level shortward of $\sim$30 \um
are mostly from the star and any unresolved inner warm component. This
is consistent with the fact that the observed spectrum between 10 and 13
\um agrees well with the expected stellar photosphere. The excess
spectrum (after stellar photospheric subtraction) is also shown in the
lower panel for both systems in Figure \ref{irs_vega_fom}.

\subsection{{\it Herschel} PACS Imaging}

\begin{figure*}
 \figurenum{3}
 \label{pacs70_images}
 \plotone{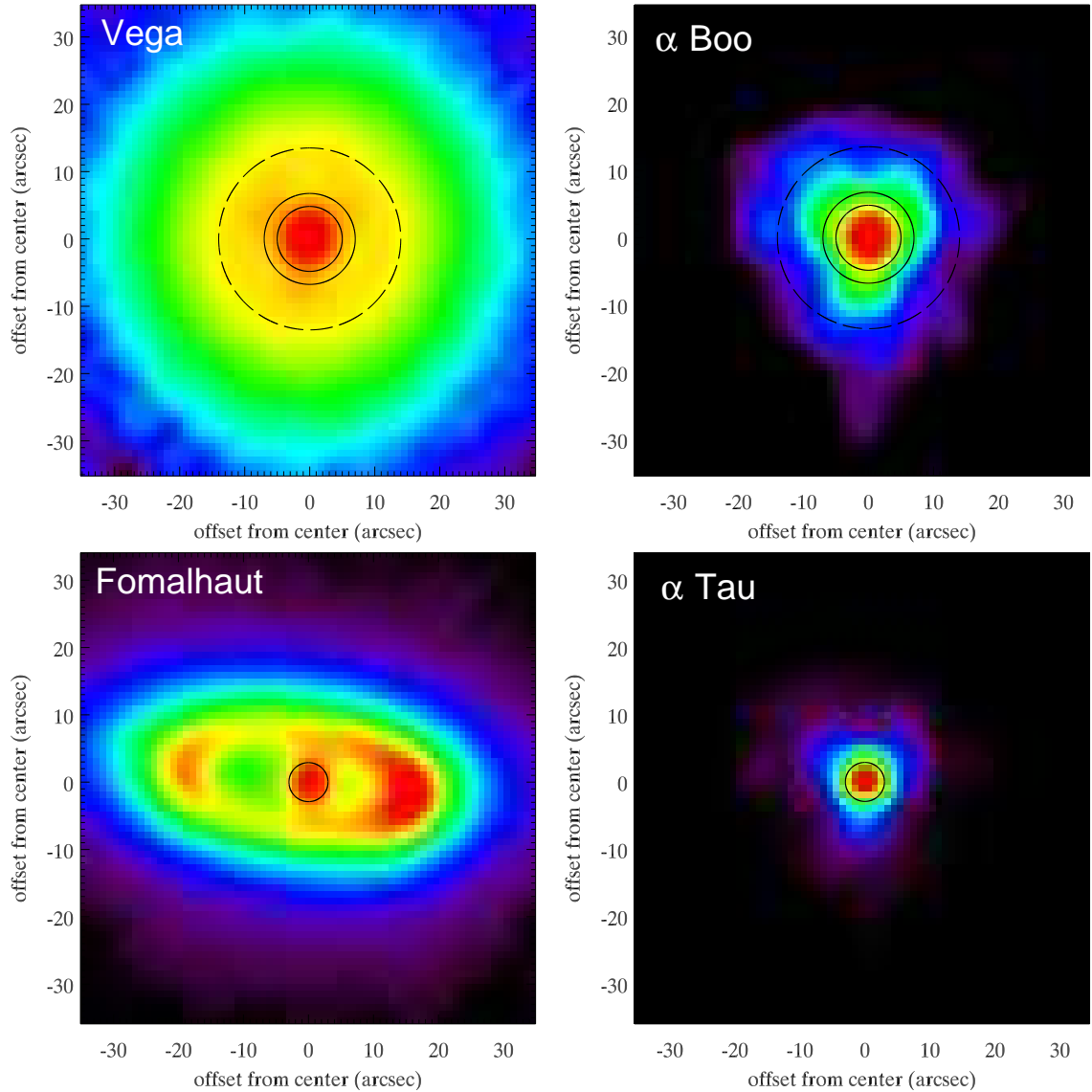}
 \caption{PACS 70 \um images of Vega and Fomalhaut along with their
reference PSF stars, $\alpha$ Boo, and $\alpha$ Tau. All images are
shown in the same angular scale and in the PACS array orientation, i.e.,
the sub-structures of the PSF are in the same orientation in all four
images. The dynamic range of display is from the peak value to 1\% of
the peak value. The color scheme is in logarithmic scale for Vega
and $\alpha$ Boo, but in squared root scale for Fomalhaut and $\alpha$
Tau for clarity. In the Vega and $\alpha$ Boo images, the two solid
circles mark radii of 5\arcsec\ and 7\arcsec\ while the dashed circle
marks a radius of 14\arcsec\ (representing the cold planetesimal
ring). In the Fomalhaut and $\alpha$ Tau images, the solid circles
mark a radius of 3\arcsec.}
\end{figure*}

{\it Herschel} PACS 70 and 160 \um data for Vega and Fomalhaut were obtained
by the GT program and published in an early reduction by
\citet{sibthorpe10} and \citet{acke12}, respectively. We retrieved the
archival data and reduced them with the {\it Herschel} Interactive
Processing Environment (HIPE, V9.0 user release, \citealt{ott10}). We
applied the standard processing steps up to the level 1 stage.  During
this process we applied 2nd level deglitching to remove outliers with
``timeordered'' option and 20$\sigma$ threshold.\footnote{Details
about the timeordered option can be found in the HIPS documentation
under the PACS photometry data
reduction, http://herschel.esac.esa.int/hcss-doc-9.0/} This is very
effective for data with high levels of coverage.  After producing
level 1 data, we selected the science frames from the time-line by
applying spacecraft-speed selection criteria (between 8\arcsec
s$^{-1}$ and 12\arcsec s$^{-1}$ for the slow scan, and 18\arcsec
s$^{-1}$ and 22\arcsec s$^{-1}$for the median scan). The final level 2
mosaics were generated using highpass filtering with the script
``photProject'' and a pixel scale of 1\arcsec\ at 70 \um and
2\arcsec\ at 160 \mm. To avoid flux loss in the highpass filtering
process, we applied a circular mask of 60\arcsec~radius centered on
the position of the target.  Since the PACS data on Vega and Fomalhaut
were obtained with different scan rates (10\arcsec s$^{-1}$ and
20\arcsec s$^{-1}$, respectively), PSF observations matched to the
observing parameters should be used for comparison. We used PACS data
on $\alpha$ Boo (ObsId 1342247634 and 1342247635) and $\alpha$ Tau
(ObsId 1342214211 and 1342214212) as our PSF reference and reduced
them using the same reduction procedure including the masking
radius. These two stars are ones of the PACS primary calibrators where
their fluxes and PSF behaviors are characterized by
\citet{muller11}. We have made sure these PSF stars have the consistent 
encircled energy fraction as a function of circular aperture radius
derived by the PACS point-source calibration \citep{muller11}. 
To illustrate the major
features seen in the PACS images, we show the final 70 \um mosaics of
Vega and Fomalhaut in Figure \ref{pacs70_images} along with the
comparison PSFs.

\section{Photospheric Properties and the Identification of the Central Unresolved Disk}

In this section, we first review all ancillary photometry to establish
the fidelity of photospheric spectrum for Vega and Fomalhaut. We then
assess whether the excess emission detected in the IRS spectrum is
also detected in the resolved {\it Herschel} PACS 70 and 160 \um
images. Since the stellar photospheres dominate the emission in the
central part of the resolved images at 70 \um where the systems are
best resolved, cares must be taken in estimating the contribution of a
central dust component without involving any further assumption of
modeling. We do this in two ways: (1) with photometry using small
apertures that exclude spatially extended emission and (2) with PSF
subtraction using reference stars to minimize the residuals at the
star position. The first approach provides an estimate of the maximum
error on the central component due to contamination from other dust
emission located in a more spatially extended distribution. The second
approach provides a more accurate estimate of the flux of an
unresolved source. In both systems, the PACS 160 \um images provide
only upper limits because the lower angular resolution at this longer
wavelength makes it difficult to spatially differentiate the
components.

\subsection{Photospheric Determination Using Ancillary Photometry}
\label{stellar}

The emission of the stellar photosphere from optical to mid-IR
($\sim$8 \mm) was determined in a number of steps. Most infrared
photometry measurements (like {\it Spitzer}/IRAC and {\it Akari}) are
referred to Vega, although Vega is unsatisfactory as a standard (i.e.,
fast-rotating, infrared excess). However, these missions also measured
Sirius using the same technique, so we have taken the measurements
with nominal uncertainties of 0.01 mag directly compared to Sirius in
terms of magnitude differences. Sirius (A1V) is very similar in
spectral type to Vega (A0V) and Fomalhaut (A4V), and is well behaved
in the infrared with no evidence for an infrared excess
\citep{price04}.

All three stars are severely saturated in the Two Micron All Sky Survey (2MASS) data. 
Therefore, for accurate measurements at 2.2 \mm, we used data from the DIRBE
instrument on {\it COBE}. The reduction of these data is described by
\citet{price10} for analyzing stellar variability. Due to the large
DIRBE beam (42\arcmin$\times$42\arcmin), the contribution from stars
in the field surrounding the target of interest needs to be
removed. We evaluated this effect using 2MASS data, making Sirius
fainter by 0.014 mag and Fomalhaut fainter by 0.003 mag while the
contribution in the Vega field is negligible. Both Vega and Fomalhaut
are reported to have $K$-band excess at 1.29\%$\pm$0.19\%
\citep{absil06} and 0.88\%$\pm$0.12\% \citep{absil09} above the
photosphere using interferometry. After accounting for these $K$-band
excesses of Vega and Fomalhaut, the $K$-band magnitude difference for
photospheres between Fomalhaut and Sirius is 2.35 mag, and 1.39 mag
between Vega and Sirius. For measurements in the IRAC bands, we
adopted the results from \citet{marengo09} who used PSF fitting
technique to recover accurate photometry for saturated sources. The
magnitude differences between Fomalhaut and Sirius are 2.36 mag, 2.36
mag, 2.36 mag, and 2.33 mag at 3.6, 4.5, 5.8, and 8 \mm,
respectively (to be discussed further by Espinoza et al.\ in
preparation). The magnitude differences between Vega and Sirius are
1.38 mag, 1.38 mag, 1.38 mag, and 1.35 mag at 3.6, 4.5, 5.8, and 8
\mm, respectively. Furthermore, {\it Midcourse Space Experiment} 
({\it MSX\ }) also measured both Vega and Sirius.  The
measured 8 \um flux of Vega is lower by 1\% compared to the {\it MSX}
predicted flux based on the measurement of Sirius \citep{price04}. We
adopt a magnitude difference of 1.36 mag at 8 \um between Vega and
Sirius.

With the nominal uncertainty for these measurements (0.01 mag), there
is no evidence for excesses from 2.2 to 8 \um (after excluding the
$K$-band excesses from interferometric measurements) for both
systems. We used appropriate Kurucz atmospheric models for
photospheric predictions in other wavelengths by normalizing the model
fluxes from 2.2 to 8 \um to the measured photometry that has been
transferred to the absolute calibration scale proposed by
\citet{rieke08}. The parameters in the Kurucz model are $T_{eff}=9500$
K and log~$g=$4.0 for Vega, and $T_{eff}=8750$ K and log~$g=$4.0 for
Fomalhaut. Given the accurate {\it Hipparcos} parallax measurements
\citep{vanleeuwen07}, the integrated luminosity is 17.45 $L_{\sun}$
for Fomalhaut and 56.05 $L_{\sun}$ for Vega (viewing from pole-on ).

Using these normalized Kurucz models, we can then predict the
photospheric level at the wavelengths of interest and determine the
excess near the stars by subtracting off the stellar contribution from
the IRS spectra presented in Section \ref{irsobs}. Note that the
Kurucz models in the infrared wavelengths are basically in the
Rayleigh-Jeans regime.

\subsection{PACS Measurements  for the Vega Central Source} 
\label{pacs_vega}

\begin{figure} 
 \figurenum{4}
 \label{pacs_profile_vega} 
 \plotone{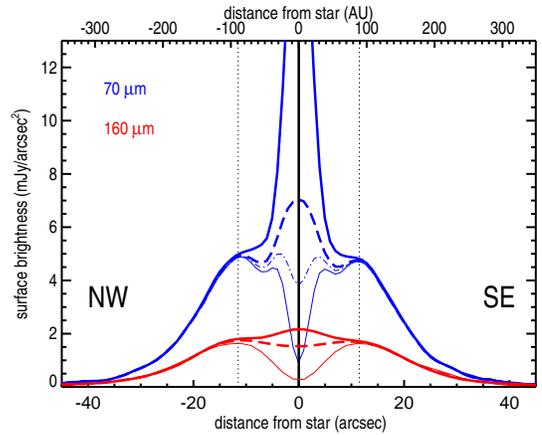}
 \caption{Surface brightness cuts for the Vega system. These cuts were
centered at the star position and represent the mean value over a
width of 5 pixels along the PA of 120\arcdeg (along the horizontal
axis in this image). The profiles at 70 \um are shown in blue color
while the ones at 160 \um in red. The original cuts (without PSF
subtraction) are shown in thick solid lines, and the long-dashed lines
represent the cuts after the subtraction of photosphere. Other thin
lines represent the cuts made with different levels of subtractions
(scaling of the PSF). For details see Section \ref{pacs_vega}. }
\end{figure}

The Vega system is viewed pole-on, and its disk has been resolved at
various wavelengths previously
\citep{holland98,heinrichsen98,su05,marsh06,sibthorpe10}. As shown in
Figure \ref{pacs70_images}, the PACS 70 \um image appears to be
centrally peaked with a smoothed, axis-symmetric extended halo.  The
size of the Vega cold disk observed in the higher resolution {\it
Herschel} data \citep{sibthorpe10} agrees with the one observed in the
{\it Spitzer} data \citep{su05}.  The stellar photosphere is about
0.81 Jy at 70 \mm, consistent with the centrally peaked morphology
seen at that wavelength. The FWHM of the central source is
5\farcs6$\times$5\farcs3 measured on a field of
21\arcsec$\times$21\arcsec\ centered on the star, which is slightly
more extended than the measured FWHM of the PSF star, $\alpha$ Boo
(5\farcs5$\times$5\farcs2).  To minimize the influence of the cold
ring (peaked at radius of $\sim$11\arcsec--14\arcsec, see Figure
\ref{pacs_profile_vega}) in estimating the photometry of the central
source, small aperture sizes with appropriate aperture corrections
should be used. On the other hand, the aperture has to be large enough
to contain most of the flux and to minimize the centroid
uncertainty. We tried several aperture settings from 3\arcsec\ to
5\farcs5 (FWHM) with and without a sky annulus and used the PSF star
$\alpha$ Boo to derive the values of aperture correction. The
resultant fluxes range from 1.00 Jy to 1.62 Jy with a median value of
1.01 Jy (using an aperture of 5\arcsec\ and sky annulus between
5\arcsec and 7\arcsec). At 160 \mm, the beam (FWHM) is
11\farcs6$\times$10\farcs1 (measured from $\alpha$ Boo), making it
very difficult to estimate the photometry of the central source alone
without significant contamination from the flux of the cold ring. A
5\arcsec\ aperture without sky annulus gives a total flux of 0.56 Jy,
which should be considered as an upper limit since it contains some
fraction of the cold ring contribution. The photosphere of Vega is
0.15 Jy at 160 \mm. Therefore, the excess flux at the star position,
based on small aperture photometry, is 0.2 Jy at 70 \um and $<$0.4 Jy
at 160 \mm.

To illustrate our PSF-subtraction approach, cuts were made along PA
of 120\arcdeg~(corresponding to the horizontal axis of the displayed
image) of the Vega system with the NW side on the negative side of the
$x$-axis. These cuts, shown in Figure \ref{pacs_profile_vega}, were
centered at the stellar position at 70 \mm, and represented the mean
value over a width of 5 pixels (i.e., 5\arcsec~and 10\arcsec~at 70 and
160 \mm, respectively). No significant asymmetry in disk brightness
nor center offset were seen in these cuts. After subtracting a
photospheric scaled $\alpha$ Boo PSF, an additional point-like source
is clearly present in the resultant cut (blue dashed line in Figure
\ref{pacs_profile_vega}) at 70 \mm. The maximum of the point-source
contribution (star and unresolved disk) can be estimated by forcing
the peak values (single pixel) matched between the Vega and $\alpha$
Boo data. A flux density of 1.19 Jy for such a PSF subtraction is
required and its resultant profile is shown as the thin solid blue
line in Figure \ref{pacs_profile_vega}, suggesting a maximum
brightness of $<$0.38 Jy for this unresolved disk at 70 \mm. Using the
flux obtained with the small aperture photometry (0.2 Jy for the
unresolved disk), the resultant profile is shown as blue dotted-dashed
line in Figure \ref{pacs_profile_vega}, relatively flat in the central
$\pm$5\arcsec~region as expected. Therefore, the brightness of the
unresolved disk in the Vega system is $\sim$0.2 Jy with a maximum
value $<$0.38 at 70 \mm. Due to the uncertainty of the PSF (as much as
10\% at 70 \mm; \citealt{kennedy12}) and that of flux calibration, we
simply assume a conservative lower-bound error of 20\% (i.e., a
minimum flux of 0.16 Jy at 70 \um for the central unresolved disk). At
160 \mm, the maximum scaled PSF is $\sim$0.45 Jy by forcing the peak
value to zero, suggesting a maximum brightness of $<$0.3 Jy for this
unresolved disk component.

\begin{figure} 
 \figurenum{5}
 \label{pacs_profile_fom} 
 \plotone{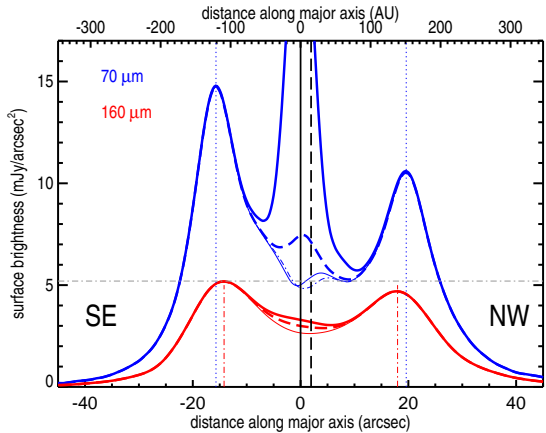}
 \caption{Surface brightness cuts along the major axis of the
Fomalhaut system. These cuts were centered at the star position
(marked as the solid vertical line) and represent the mean value over
a width of 4 pixels along the minor-axis. Line styles and colors are
similar to those used in Figure \ref{pacs_profile_vega}. The dotted
and dotted-dash vertical lines mark the peak positions of the ansae at
70 and 160 \mm, respectively, while the long-dashed vertical line represents
the mid-point of the cold ring. }

\end{figure}

\subsection{PACS Measurements  for the Fomalhaut Central Source} 
\label{pacs_fom}

Since the Fomalhaut debris system is inclined by 67\arcdeg, it is very
difficult to separate the point source from the cold ring even using
very small apertures.  One can still estimate the maximum brightness of
the unresolved component by assuming there is no contamination from
other sources inside a very small aperture. For an aperture radius of
3\arcsec, the encircled flux is 0.95 Jy at 70 \um and 0.77 Jy at 160
\um after applying appropriate aperture corrections. Taking out the
contribution of the stellar photosphere (0.368 Jy and 0.07 Jy at 70
and 160 \mm, respectively, based on our stellar photospheric model),
the unresolved disk component has a maximum flux of $<$0.58 Jy and
$<$0.7 Jy at 70 and 160 \mm, respectively.

Similar to Vega, cuts were made along the major-axis of the disk and
shown in Figure~\ref{pacs_profile_fom}. These cuts were centered at
the stellar position at 70 \mm, and represented the mean value over a
width of 4 pixels (i.e., 4\arcsec~and 8\arcsec~at 70 and 160 \mm,
respectively). The center of the outer cold ring (the dashed, vertical
line in Figure \ref{pacs_profile_fom}) was estimated by the mid-point
of the two bright peaks (marked as dotted, vertical lines for 70 \um
and dot-dashed, vertical lines for 160 \um in Figure
\ref{pacs_profile_fom}) at both wavelengths. The 160 \um ansae peak
closer to the ring center than the ones at 70 \mm, which was first
reported by \citet{acke12} and suggested to be due to blurring in the
large PSF at 160 \mm.  Although the peak positions at 70 and 160 \um
are not the same, the cold ring centers at the same position at both
70 and 160 \mm. We used a PSF star, $\alpha$ Tau, for photospheric
subtraction by scaling it to match the expected photospheres (0.368 Jy
and 0.07 Jy at 70 and 160 \mm). The photosphere-subtracted cuts are
shown in long-dashed lines in Figure~\ref{pacs_profile_fom}. At 70
\mm, it is clear that an additional source of emission is present at
the center of the disk. We tried two different PSF subtractions to
estimate the brightness of this additional component. First, we
arbitrarily increased the scaling of $\alpha$ Tau, fixed in the
stellar photosphere position, until the central region (from
$-$5\arcsec~to $+$5\arcsec~) has a relatively flat distribution in the
cut. The resultant cut is shown as a thin solid blue line with an
additional scale of 0.17 Jy. Second, we subtract a second $\alpha$
Tau PSF by adjusting both the position and the scaling until the
resultant cut is relatively flat (dot-dashed line in Figure
\ref{pacs_profile_fom}). The second method gives a scale of 0.165 Jy
for this additional source. Combining both methods, we conclude that
the central unresolved component is 0.17 Jy at 70 \mm. There is no
easy way to estimate the error associated with this number since it
depends on the detailed structures of various components, we simply
assume a 20\% error for further analysis. At 160 \mm, it is not
obvious that an additional source is required in the system. We
estimated the upper limit of such a component by increasing the
scaling factor of $\alpha$ Tau fixed at the star position (any
potential offset between the star and this inner component is washed
out by the large beam size at 160 \mm). An upper limit of 0.08 Jy at
160 \um was inferred.

In summary, the central unresolved disk component in the Fomalhaut
system is about 0.17 Jy at 70 \um with upper- and lower-bound fluxes
of 0.58 Jy and 0.136 Jy (assuming 20\% uncertainty), and $<$0.08 Jy at
160 \mm, based on the analysis of the data only (without invoking any
assumption of modeling). Our values are consistent with the best-fit
model presented in \citet{acke12}. Their model estimates that 
the fluxes of
the central point source (star + unresolved disk component) are 0.54
Jy (5\% of the total flux) and 0.124 Jy (2\% of the total flux) at 70
and 160 \mm, respectively, implying that the unresolved disk accounts
for flux of 0.172 Jy at 70 \um and 0.054 Jy at 160 \mm.

\section{Analysis} 
\label{analysis}

\subsection{Unresolved Warm Excess in the Vega System}

\begin{figure}
 \figurenum{6}
 \label{asteroid_vega}
 \plotone{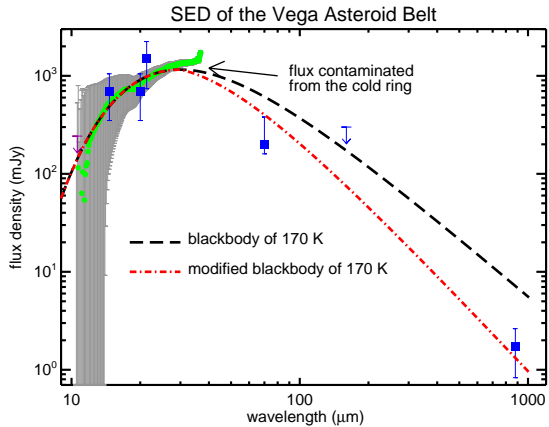}
 \caption{Excess SED of the inner warm component in Vega. The
photospheric subtracted IRS spectrum is shown as green dots with
errors (including the 2\% errors from the photosphere) shown as gray area; 
other IR and submillimeter excess photometry are
shown as blue squares. For comparison, blackbody curves of 170 K are
also shown with $B_{\lambda}$ as the long-dashed line and
$\lambda^{-0.5}{}B_{\lambda}$ as the dotted-dash line. The 10.6 \um
BLINC/MMT 1$\sigma$ upper limit is also shown.}
\end{figure}

The Vega {\it Spitzer} 24 \um observation \citep{su05} was severely
saturated, making it difficult to constrain the brightness of this
unresolved component without invoking some modeling
assumptions. Therefore, we seek other relevant measurements in the
mid-infrared to validate the excess levels seen in the IRS
spectrum. \citet{tokunaga84} measured a handful of nearby A-type stars
at 20 \um using the IRTF bolometer with a beam size of 5\arcsec\ to
define the 20 \um magnitude system (relative to Vega). Only two stars,
$\alpha$ CMa (Sirius) and $\gamma$~UMa, in his list do not have a 24
\um excess. We used the color $V-$[20] of these two stars to extract
the excess of Vega at 20 \mm. We adopt $V$ of $-$1.40 and 2.40 mag for
$\alpha$ CMa and $\gamma$~UMa, suggesting a photosphere color $V-$[20]
of $-$0.04 mag in Tokunaga's system. Based on the observed $V-$[20]
color of 0.03, the excess of Vega is $\sim$7\% above the photosphere
(i.e., 0.7 Jy). It is difficult to assess the errors associated with
the 20 \um measurement; therefore, we simply assume a maximum 50\%
error at this wavelength.  Another source of mid-IR measurements for
Vega comes from {\it MSX} photometry where the excesses (relative to Sirius)
at 14.65 and 21.34 \um have been reported by \citet{price04} to be 4\%
and 17\% above the photosphere (i.e., excesses of 0.7 Jy at 14.65 \um
and 1.5 Jy at 21.34 \mm), respectively. The contamination from the
cold ring needs to be taken into account given the large beam size of
{\it MSX} (a resolution of $\sim$20\arcsec). Assuming a typical dust
temperature of 60 K for the cold ring and normalizing the cold ring
flux observed in the far-infrared \citep{su05}, the flux contamination
from the cold ring is less than 0.1\% of the photosphere at 14.65 \mm,
and $\sim$4\% of the photosphere at 21.34 \mm. The nominal error for
the {\it MSX} measurements is $\sim$1.5\% \citep{rieke08}, which includes
the errors from stellar photospheric prediction. Including the
possible contamination from the cold ring, the final errors are 0.31
Jy and 0.38 Jy at 1-$\sigma$ for the excesses at 14.65 and 21.34
\mm. The Vega system was recently observed by the Submillimeter Array
(SMA) at 880 \um \citep{hughes12}. Interestingly, the spatially
unresolved 880 \um flux within 5\arcsec\ of the Vega photosphere lies
slightly above the predicted photospheric value (see Figure
\ref{asteroid_vega}), although only at the less than 2$\sigma$
level. Combining with our estimates from the PACS images and the IRS
spectrum, the excess SED of the central component in the Vega system
is shown in Figure \ref{asteroid_vega}. We note that \citet{liu04}
also report an upper limit ($\sim$250 mJy) of 0.7\% (1$\sigma$) of the
photospheric level at 10.6 \mm, using nulling interferometry
(BLINC/MMT) that probes a region less than 1\farcs5 (12 AU) from the
star.

The spectral shape ($\lambda<$30 \mm) of the inner component is well
presented by a blackbody of 170 K, just above the temperature at which
water ice sublimates in vacuum. This temperature corresponds to a
distance ($\sim$2.7 AU) in the middle of our asteroid belt in the
solar system, but $\sim$14 AU in the Vega system (using an average
stellar luminosity of 37 $L_{\sun}$ viewed by the dust along the
equator of the star; \citealt{aufdenberg06}) assuming blackbody-like
emitters.

\subsection{Unresolved Warm Excess in the Fomalhaut System}

\begin{figure}
 \figurenum{7}
 \label{asteroid_fom}
 \plotone{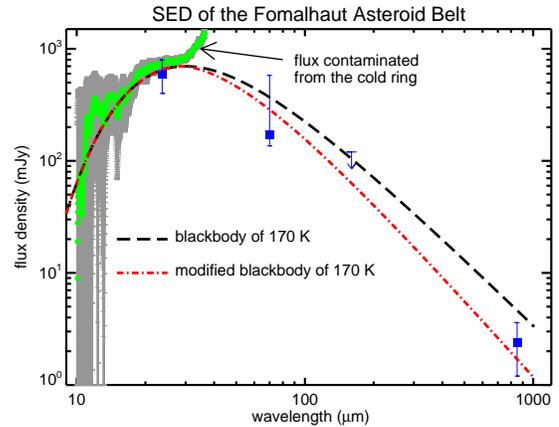}
 \caption{Excess SED of the inner warm component in Fomalhaut.
Symbols and lines are similar to what were used in
Figure \ref{asteroid_vega}. The upper error bar at 70 \um represents the
(unlikely) maximum value estimated from small aperture photometry and
the lower error bar is 20\% lower than the estimate from PSF
subtraction. Other error bars are shown as 1-$\sigma$. For comparison,
blackbody curves of 170 K are also shown with $B_{\lambda}$ as the
long-dashed line and $\lambda^{-0.3}{}B_{\lambda}$ as the dotted-dash
line.}
\end{figure}

The existence of a close (unresolved) warm component around the
Fomalhaut disk was first suggested by the {\it Spitzer} 24 \um
observation where an additional point-like source (0.6$\pm$0.2 Jy)
along with the expected stellar photosphere is required to fit the
resolved disk image at 24 \um \citep{stapelfeldt04}.  The central
component of the Fomalhaut system was also tentatively detected by the
ALMA observation at 850 \um \citep{boley12}. A flux density of
$\sim$4.4 mJy at the star position was estimated after the primary
beam correction, which is quite uncertain at the star position
because the ALMA observation was centered at Fomalhaut b. The expected
photosphere at 850 \um is $\sim$2 mJy (not 3 mJy as stated in
\citealt{boley12}). Therefore, the excess at 850 \um could be as high as
2.4 mJy. We simply took this at a face value, and assumed a 50\% error
(as maximal). The excess SED for the Fomalhaut central unresolved
component is shown in Figure \ref{asteroid_fom} along with the excess
spectrum measured by IRS.

We note that \citet{acke12} suggested ionized gas (free-free emission
from a hot stellar wind) being responsible for the unresolved central
excess from $K$ band to 70 \mm, and derived $F_{\nu}\propto
\nu^{0.8\pm0.1}$ power law. With their derived power-law, the excess
at 850 \um (350 GHz) is expected to be 18--23 mJy, which is clearly
not consistent with the ALMA observation. In addition, this power-law
spectrum is inconsistent with the 2.2--5.8 \um colors discussed in Section
\ref{stellar}. Instead, we suggest that the excess emission detected
longward of $\sim$13 \um arises from thermal dust emission. Similar to the
warm component in the Vega system, the spectral shape ($\lambda <$30
\mm) of the inner component in Fomalhaut is also well representative
by a blackbody of 170 K, suggesting a radial distance of $\sim$11 AU
from the star for blackbody-like grains.

\subsection{Asteroid-belt Analogs}
\label{asteroid-analog}

The warm excesses detected in both systems are restricted to the
vicinity of the star; i.e., unresolved at various wavelengths. The
amount of warm excess emission is derived either from resolved images
where the central component is (mostly) separated from the cold ring
or from the spectra taken centered at the star position. Given the
sizes of the IRS slits, some amount of the flux from the cold rings is
expected to fall into the slit, especially at wavelengths longer than
$\sim$30 \mm. It is difficult to estimate the exact flux contamination
without further modeling because it depends on the geometry of the
cold ring at different wavelengths. One can estimate the maximum flux
contamination in the worst scenario case by assuming the warm and cold
components are spatially coincident. Therefore, the maximum flux
contamination is $\lesssim$1\% of the photosphere for $\lambda <$20
\um and could be up to 50\% at 33 \um given the typical dust
temperature of the cold ring ($\sim$70--50 K). Thus the derived dust
temperature ($\sim$170 K in both systems) based on the data shortward
of 30 \um has very little contamination from the cold ring. In other
words, the excess arises from material less than 3\arcsec\ (unresolved
with an FWHM$\sim$6\arcsec) from the star in both cases. Similar to the
definition of habitable zones around stars, debris disk structures
should be identified and characterized in terms of dust temperatures
rather than physical distances so that the heating power of different
spectral type of stars is taken into account and common features in
disks can be discussed and compared directly. The characteristic
temperature of $\sim$170 K for the excesses suggests they are
asteroid-belt analogs. It corresponds to a distance of 2\arcsec
($\sim$14 AU) from Vega (using the lower luminosity viewed by the dust
in the equatorial direction) or of 1\farcs5 ($\sim$11 AU) from
Fomalhaut assuming blackbody radiators. The resultant location of the
asteroid belt given a dust temperature depends greatly on grain
properties. Using astronomical silicates \citep{laor93}, the dust can
be as close as 1\farcs3 (10 AU) from Vega, and 1\arcsec~(8 AU) from
Fomalhaut using grains with a radius of 10 \mm. Furthermore, the IRAM
PdBI observations at 1.3 mm by \citet{pietu11} do not find any excess
within 2\farcs5 while the SMA data by \citet{hughes12} suggest a
tentative excess within 5\arcsec\ from the star, suggesting that the
warm excess around Vega likely arises from emission outside a radius
of 1\farcs3 (10 AU) from the star. The fractional luminosity is
7$\times10^{-6}$ for the Vega warm component and 2$\times10^{-5}$ for
the Fomalhaut warm component, much more luminous than our current
zodiacal cloud ($10^{-8}-10^{-7}$; \citealt{dermott02}).

The origin of the warm excess in the vicinity of the star is
unknown. One hypothesis is that the warm excess originates from dust
grains in the cold belt, which are transported inward by P-R and/or
stellar wind drag, as suggested for the outer warm belt of $\epsilon$
Eri debris system by \citet{reidemeister11}. In the Fomalhaut system,
the presence of an inner disk, extending inward up to $\sim$35 AU
(4\farcs5) and composed of grains dragged-in from the cold
planetesimal belt, has been suggested by \citet{acke12} to explain the
PACS 70 \um profile. Conventionally, a PR-transported flow has a
constant surface density and is expected to extend to the star,
resulting in a surface brightness profile that is centrally peaked at
the star. However, \citet{acke12} find that the inner disk in
Fomalhaut has a clear truncation (outside the unresolved warm
component). Therefore, neither the warm nor the hot excess observed in
the Fomalhaut system is unlikely to have arisen from the inward
transport of grains by P-R drag.  On the other hand, we cannot rule
out such a possibility (either solely or partially due to the
dragged-in grains) in the Vega system (it is centrally peaked at 70
\mm) without detailed modeling for the whole system at multiple
wavelengths (K.\ Y.\ L.\ Su et al.\ in preparation). Nevertheless, the
observed dust temperature (170 K) of the warm excess suggests this
component does not extend all the way to the star; instead, an inner
truncation ($>$1\arcsec) is required.

Furthermore, the tentative detection of the central component at
submillimeter wavelengths in both systems argues for the presence of
large grains, suggesting the warm emission likely arises from a full
spectrum of particle sizes similar to the cold planetesimal belt,
i.e., a form of planetesimal belt like our asteroid belt. The
wavelength-dependent power index at long wavelengths ($F_{\nu}\sim
\lambda^{-l}$) is a measure of the grain size distribution in a
collision-dominated debris disk \citep{wyatt02,gaspar12}. The cold
planetesimal belts observed in bright debris disks have much steeper
slopes compared to Rayleigh-Jeans slope (e.g.,
\citealt{gaspar12}). Similar behavior is seen in the warm component in
the Vega and Fomalhaut systems (see Figures\ \ref{asteroid_vega} and
\ref{asteroid_fom}), implying the presence of large grains. 
Limited by the uncertainties in the current
observations, the exact slope of the warm component at long
wavelengths cannot be determined accurately. Future high-resolution
observations in the submillimeter and millimeter will help shed light
on the nature of the warm excesses. 

One interesting note on the nature of these warm excesses is that the
observed levels of dust ($f_d\sim 7\times10^{-6}$ for Vega and
$f_d\sim 2\times10^{-5}$ for Fomalhaut) are consistent with being the
in situ, steady-state collisional evolution of large parent bodies
for the lifetime of the stars ($\sim$400 Myr). The maximum fraction
luminosity ($f_{max}$) for such a system can be estimated using
Equation (18) from \citet{wyatt07} where $f_{max}$ is on the order of
(2--3)$\times10^{-5}$ for both systems assuming an asteroid belt at
$\sim$10 AU with a width of $\sim$1 AU and consisted of $\sim$10 km
(diameter) planetesimals in collisional cascades. Due to the
uncertainties of some parameters in the model, \citet{wyatt07} suggest
that a transient event producing the observed dust is only required
when $f_{obs} \gg 10^3 f_{max}$. Thus, the observed dust in the warm
component is consistent with it being generated through collisional
grinding in an asteroid belt in both systems.

\section{Discussion}

Vega and Fomalhaut really live up to their names as debris disk twins:
both possess a hot 2 \um excess revealed through interferometric
observations, a cold ($\sim$50 K) belt analogous to our Kuiper belt
that has been in the spotlight of space infrared facilities, and a
warm ($\sim$170 K) belt analogous to our asteroid belt. This
warm-and-cold-belt debris architecture has also been identified in HR
8799 \citep{su09} and $\epsilon$ Eri \citep{backman09}. In Section
\ref{origin}, we first discuss the possible origin of these two-belt
systems in terms of their formation and evolution. Taken together with
the known properties of planets and mass limits from non-detection,
we speculate that the large gap between the two belts is likely to be
sustained by multiple (low-mass) planets. We then review the properties of
2 \um excess in Section \ref{hotexcess}, and suggest an alternative
explanation for the hot dust component.

\subsection{Origin of the Two-belt Systems}
\label{origin}

In our solar system, the locations of the minor bodies that failed to
form planets are elegantly arranged and sculpted by planetary
perturbations over the course of 4.5 Gyr evolution.  The inner edge of
the Kuiper belt's dusty disk is thought to be maintained by massive
planets \citep{liou99}, whereas the more tenuous asteroid-belt dust
(i.e., zodiacal cloud) has a structure also determined by
gravitational perturbations via both the giant and terrestrial planets
\citep{dermott94,murray98}.  It has been suggested that the dominant
source of dusty debris inside $\sim$5 AU in our solar system results
from the breakup of asteroids (e.g., \citealt{dermott02}).  However, a
recent dynamical model incorporating multiple sources (asteroids and
short- and long-period comets) by \citet{nesvorny10} suggests that
particles originated from Jupiter-family (short-period) comets
dominate the mid-infrared emission in the zodiacal cloud while the
contribution of asteroidal dust is $<$10\%. The origin of our own
zodiacal cloud is still a matter of considerable debate, making
understanding of the warm dust around nearby stars (i.e., exo-zodi)
even more valuable.

\begin{figure}
 \figurenum{8}
 \label{amazing4}
 \plotone{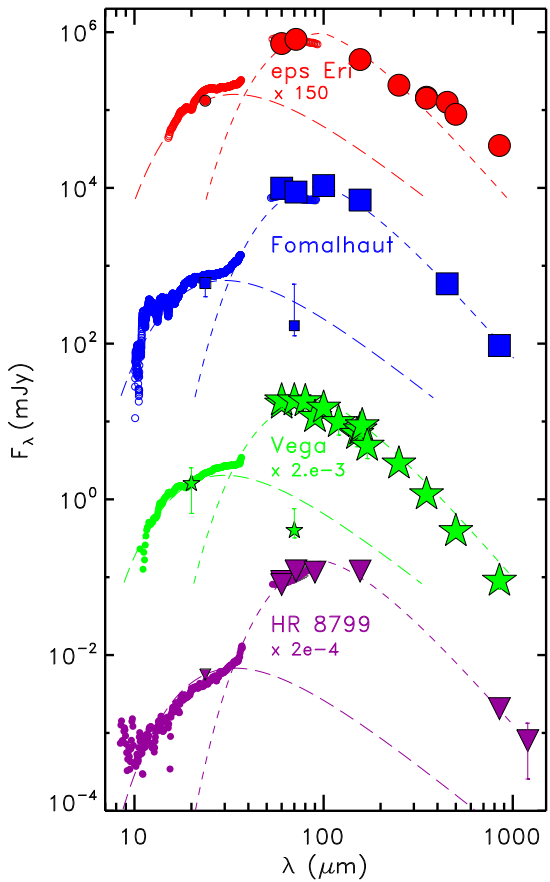}
 \caption{Excess SEDs for $\epsilon$ Eri, Fomalhaut, Vega and HR 8799
where measurements are shown in various symbols with colors
differentiating the objects and large size symbols being the
integrated photometry while smaller ones representing spectroscopic
measurements. The photometry of the warm components is also shown as
the smaller symbol. The excess of each system is well represented by
two (warm as long-dashed line and cold as dashed line) blackbody
emissions: $\epsilon$ Eri: 150 K and 50 K, Fomalhaut: 170 K and 50 K,
Vega: 170 K and 50 K, and HR 8799: 150 K and 45 K. Note that the cold
component is modeled with modified blackbody as
$\lambda^{-0.9}{}B_{\lambda}$ in order to fit the sub-millimeter
observations.}
\end{figure}

Dust locations solely based on temperatures derived from ex-solar
debris disk SEDs are ambiguous. However, based on resolved images at
multiple wavelengths, four systems -- Vega, Fomalhaut, $\epsilon$
Eri,\footnote{For the $\epsilon$ Eri disk, we refer the inner warm belt
that dominates the emission at 24 \um rather the outer warm belt that
emits prominently at 55 \um \citep{backman09}.}  and HR 8799 --
clearly have two dust belts.  The disk SEDs of these four systems are
shown together in Figure \ref{amazing4} for easy comparison. As
discussed in Section \ref{asteroid-analog}, the dust observed in the
warm components for both Vega and Fomalhaut is consistent with being
generated in a steady-state collisional cascade in a planetesimal
belt. A similar conclusion obtains for the other two systems. For HR
8799, the warm dust component has a temperature $\sim$150 K and
fractional luminosity $f_d\sim2\times10^{-5}$ (\citealt{su09}); the
latter is much less than the maximum fractional luminosity for steady
state collisional dust production, $f_{max}\approx3\times10^{-4}$,
estimated for a belt at 8 AU at HR 8799's age of $\sim$30 Myr.  For
the $\epsilon$ Eri system, the warm component has
$f_d\sim3\times10^{-5}$ and temperature of $\sim$150 K
(\citealt{backman09}); the latter is compatible with
$f_{max}\approx8\times10^{-6}$ for a belt at 3 AU at an age of
$\sim$800 Myr.  Based on the similarity of the observed SEDs and
resolved disk structures, it is very likely that the warm components
in these four systems originate in well-separated planetesimal belts
(remnants of planet formation), resembling the configuration of our
own solar system which has two left-over planetesimal belts: an
asteroid belt near the water-frost line, between Mars and Jupiter,
with a characteristic temperature of $\sim$170--150 K, and a Kuiper
belt at $\sim$30--55 AU with dust emission peaked at $\sim$50 K.

There are two aspects of the two-belt systems that must be
explained. The first one is related to how these two separate belts
were created, and the other one addresses how the system maintains
such a large gap without dust filling in it. Mechanisms to explain
both the formation and evolution of the two belt systems are required
to fully account for the observed pattern. Planetesimal belts can be
expected in regions where there was not enough material or not enough
time to form a planet, because once formed, a planet would scatter or
accrete the surrounding planetesimals. Therefore, the stable locations
of left-over planetesimals are governed by where the giant planets
form and their migration history, which may include strong dynamical
instabilities. Consequently, the location of the observed dust is not
expected to show much order among systems since it greatly depends on
the numbers and positions of the giant planets and their migration
history. In addition, it is an observational fact that higher mass
stars harbor a more massive protoplanetary disk
\citep{natta00,williams11}, and more giant planets
\citep{johnson10}. Generally, one does not expect to find similarity
in planetary configurations between low-mass (i.e., low luminosity)
and high-mass (i.e., high luminosity) stars.

Among the four systems shown in Figure \ref{amazing4}, only $\epsilon$
Eri is a solar-like star and the rest are early-type stars.  The
luminosity (the dust heating power) difference ranges roughly two
orders of magnitude among them, and yet the characteristic dust
temperatures of the warm and cold belts are very similar. It is
interesting to note that many unresolved systems also have a similar
two-belt configuration in the dust distribution based on their
temperatures derived from unresolved excess emission
\citep{chen09,morales11}. The fact that the temperature of the warm
belts peaks at similar temperatures ($\sim$170--190 K) between
early-type and solar-like systems \citep{morales11}, however, suggests
temperature-sensitive mechanisms play a major role in determining
planetesimal configuration in the warm belts. Giant planets are
expected to form right outside the water-frost line where the amount
of solid material and dynamical timescales favor the formation process
\citep{kretke07}.

The young HR 8799 system has four giant planets \citep{marois08,
marois10} separating the inner and outer dust belts
\citep{su09}. These four planets are very massive ($\sim$7--10 $M_J$)
so that they dominate the dynamics in the HR 8799 system. Dust
particles spiraling inwards from the cold belt under P-R drag are
likely to be dynamically scattered and ejected by one of the planets
before they reach the inner region. Thus, the warm component in the HR
8799 system is unlikely to arise from grains generated in the cold
planetesimal belt. Comets, on plunging orbits originating from the
cold belt could possibly cross over and then break-up or sublimate to
deliver dust interior to the giant planets. However, it is difficult
to estimate the inward flux of cometary bodies inside the orbits of
the giant planets without detailed numerical simulations such as the
one done by \citet{bonsor12b}. Furthermore, the comet delivery
scenario proposed by \citet{nesvorny10} might work for our solar
system at its current age, it is not clear that this scenario is
directly applicable for younger systems (such as the ones we discussed
here) where in-situ dust generation in the younger, more massive
planetesimal belts is expected to dominate.

In the remaining three systems, Fomalhaut is the only other one that
harbors a directly detected planet, Fomalhaut b
\citep{kalas08}. Although the reality of Fomalhaut b has been
questioned because the spectrophotometry of Fomalhaut b cannot be
reconciled with models for thermal emission from a young giant planet,
a problem originally acknowledged in \citet{kalas08} and further
demonstrated by {\it Spitzer} non-detections
\citep{marengo09,janson12}, a perturbing planet is required to account
for the observed disk asymmetry \citep{kalas05,quillen06,chiang09}.
Nevertheless, recent re-analyses on the public {\it HST} data by
\citet{currie12} and \citet{galicher12} confirm the detection of
Fomalhaut b with its position found to be just interior to the ring
and comoving with the star, making it a candidate to maintain the
sharp inner boundary of the Fomalhaut cold ring. The orbit of
Fomalhaut b has not yet been fully demonstrated to be consistent with
the range of orbits required for the perturber, and its mass is also
quite uncertain due to lack of detections at other wavelengths,
ranging from $<$a few $M_J$ (from the constraint of cold belt as
detected in scattered light, \citealt{chiang09}) to a few $M_{\earth}$
(from the constraint derived from the properties of the cold belt in
the submillimeter, \citealt{boley12}). Furthermore, ground-based
high-contrast observations have also set a mass limit of $<$2 $M_J$
for any planet between $\sim$10--40 AU \citep{kenworthy09} and of
$<$10--16 $M_J$ from 3 to 10 AU \citep{kenworthy13} in the Fomalhaut
system.

The clumpy structures claimed from millimeter and submillimeter
imaging of the Vega cold Kuiper-belt analog were often taken as
signatures of gravitational perturbations by planets
\citep{holland98,wilner02,wyatt03,marsh06}. Recent observations at
higher resolution and sensitivity fail to detect the clumps, and
instead are consistent with a smooth, broad, and axisymmetric disk
\citep{pietu11,hughes12}. In any case, Vega has been the target of
several deep searches for planets through direct imaging, and strong
limits have been placed at $H$, $L'$, and $M $ bands for planets with
masses $\gtrsim 3-4$ $M_J$ between $\sim$20 and $\sim$70~AU
\citep{marois06,heinze08}.  For $\epsilon$ Eri, planets with masses
$\gtrsim$2--3 $M_J$ in radial distance of 6--35 AU are ruled out at
$H$, $L'$, and $M $ bands \citep{lafreniere07,heinze08} and {\it
Spitzer} IRAC bands \citep{marengo06,marengo09}. Using radial velocity
(RV) and astronometry techniques, a close, eccentric planet,
$\epsilon$ Eri b, with a semi-major distance of $\sim$3--4 AU was
identified by \citet{hatzes00} and \citet{benedict06}; although the
discovery of the inner warm dust belt ($\sim$3 AU) casts doubt on the
existence of a high eccentricity planet
\citep{backman09}. Furthermore, a recent analysis of all available RV
data of $\epsilon$ Eri by \citet{anglada12} found a significantly
different orbital solution and suggested that the long-term RV
variability is likely due to stellar activity cycles rather than a
putative planet. Nonetheless, $\epsilon$ Eri b, if it exists just outside
the warm belt, is a candidate to shepherd the warm dust belt
\citep{backman09}.

Although the exact locations of the warm components in these disks are
unknown (unresolved), the orbital ratios ($R_{cold}/R_{warm}$) between
the outer cold belt (mostly resolved in the submillimeter and millimeter
wavelengths) and the warm belt estimated from the dust temperature are
roughly $\gtrsim$10 in all four systems, \footnote{Vega:
$R_{warm}\sim$14 AU and $R_{cold}\sim$110 AU; Fomalhaut:
$R_{warm}\sim$10 AU and $R_{cold}\sim$140 AU; HR 8799:
$R_{warm}\sim$10 AU and $R_{cold}\sim$100 AU; $\epsilon$ Eri:
$R_{warm}\sim$3 AU and $R_{cold}\sim$35 AU.} similar to the ratio in
our own solar system (the asteroid belt at $\sim$3 AU and the Kuiper
belt at $\sim$35 AU).  A significant deficit of large grains (best
tracers for the unseen parent bodies) in the region between the belts
are evident due to the fact that high-resolution submillimeter
observations do not detect such a filled-in component \citep{boley12},
although small grains can drift inward from the cold belt
\citep{acke12}. The zone between the warm and cold belts that is
mostly free of dust is very large. If the observed dust is being
generated in both belts through collisional cascades of large parent
bodies, we need a cleaning mechanism to maintain such a large
dust-free zone.

\begin{figure}
 \figurenum{9}
 \label{chaotic}
 \epsscale{1.0} 
 \plotone{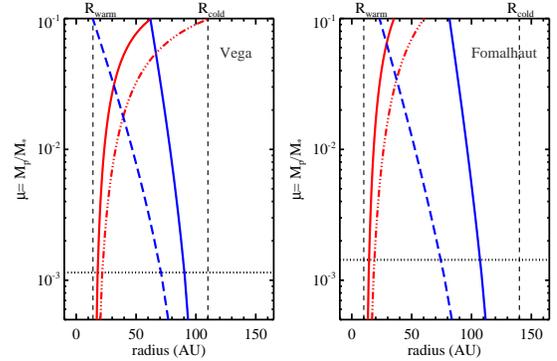}
 \caption{Mass-dependent chaotic zones for two equal-mass planets
in the Vega (left panel (a)) and Fomalhaut (right panel (b)) systems, related
to the boundaries of the inner and outer belts (shown as black,
dashed, vertical lines). In both panels, the orbital radius of the
inner planet is shown as red solid lines while the one of the outer
planet is shown as blue solid lines. In Vega, the chaotic zone width
is computed using $e=$0 for the planet's eccentricity, but for
Fomalhaut we adopt $e=$0.1. The dotted-dash lines represent the outer
boundary of the chaotic zone for the inner planet while the
long-dashed lines represent the inner boundary of the chaotic zone for
the outer planet. For two equal-mass planets, the planet-star mass
ratio, $\mu$, has to be greater than 0.015 (0.03) for these two planets
to maintain the dust-free zone in the Vega (Fomalhaut) system. The
current observational upper limit for the planet mass is shown as the
horizontal dotted lines. }

\end{figure}

The large gap between the warm and cold belts may be maintained by one
or multiple planet-mass objects in the gap. If these planets have
dynamical influence over the entire gap, they are likely to scatter
any material that is either generated in the gap or drifts into
it. The dynamical influence of a planet is given approximately by the
extent of the overlap of first-order resonances which creates an
unstable ``chaotic zone'' in the vicinity of a planetary perturber
(\citealt{wisdom80,duncan89,mustill12}).  Applying this criterion, we
find that in the Vega system, the large gap could be maintained by a
single hypothetical object with mass of a few 100 $M_J$ (i.e., a star/brown
dwarf) in a circular orbit; for Fomalhaut, a similar mass estimate of
a few 100 $M_J$ obtains for a single hypothetical object in an
eccentric orbit ($e=0.1$, based on the observed eccentricity of its
cold dust ring). One can also imagine a lower mass perturber in very
eccentric orbit being responsible for maintaining such a large gap;
the required eccentricity can be estimated by assuming that its
pericenter and apocenter are near the locations of the warm and cold
belts. We find that a large eccentricity ($e\sim$0.8) is required,
implying an object with a mass of $\sim$50 $M_J$ in both cases.  Our
simple estimates show that the gaps in both Vega and Fomalhaut are too
large to be explained by a single circular/eccentric perturber,
without contradicting the current upper limits of a few $M_J$ based on
non-detections of planetary companions in the gaps.

The minimum number of planets residing between the two belts is two --
one inner planet outside the warm dust belt to shepherd the inner
planetesimal belt, and one outer planet interior to the cold dust belt
to scatter large grains that drift inwards from the cold belt. We can
then estimate the mass of two equal-mass planets in this case. This is
illustrated in Figure \ref{chaotic} where we plot the orbital
distances of the two equal-mass planets and their associated chaotic
zone widths as a function of the mass ratio between the planet and the
host star. In the case of Vega ($M_{\ast}=2.5M_{\sun}$ and we
assume $e=0$ for the hypothetical planets), we find that two $\sim$40
$M_J$ planets are needed to maintain the dust free zone.  In the case
of Fomalhaut ($M_{\ast}=2.0M_{\sun}$ and we assume $e=0.1$ for the
hypothetical planets), we find that two $\sim$63 $M_J$ planets are
needed. Although these masses are likely overestimates because the
single-planet chaotic zone formulae do not account for the strong
secular perturbations that can extend the unstable zones in multiple
planet systems (e.g., \citealt{moro-martin10}), it is evident that
just two planets in low eccentricity orbits having mass $\sim M_J$ are
inadequate for explaining the gaps.

Our simplified approach, in estimating the masses for single planetary
perturbers and for two equal-mass perturbers for explaining the large
gaps in Vega and Fomalhaut, gives limits that are more than an order
of magnitude higher than the $\sim$few $M_J$ planet mass limits for
these systems based on non-detections of planetary companions in the
gaps.  It is, therefore, probable that multiple, lower mass planets are
responsible for clearing the gaps. We can estimate the numbers and
orbital radii of multiple, Jupiter-mass planets that would clear this
region, by assuming that they are separated such that their chaotic
zones just fill the regions between them. We find that four or five
$1M_J$ mass planets are required for the large gap in Vega and in
Fomalhaut. (This is reminiscent of the HR 8799 system in which four
planets separate the two dust belts.) Upcoming ground-based
high-contrast, direct imaging surveys using facilities like LBT/LBTI,
Gemini/GPI, VLT/SPHERE should be able to find and/or place tighter
mass limits on the planets in the large gap. Stars that have a debris
disk with two distinct dust belts separated by a large gap are
attractive targets for future searches of exoplanets.

\subsection{The Mystery of the Hot Dust }
\label{hotexcess}

The presence of hot dust in the close vicinity of both systems is
revealed by ground-based interferometric observations. For Fomalhaut,
0.88\%$\pm$0.12\% of excess emission in the $K$ band was reported by
\citet{absil09}. For Vega, a 1.29\%$\pm$0.19\% excess in the $K$ band
\citep{absil06} and 1.23\%$\pm$0.45\% excess in the H band
\citep{defrere11} were reported. Generally, these studies have ruled
out other possible sources such as low-mass companions or stellar
winds to be the cause of excess emission. \citet{mennesson11} also put
a very tight constraint on the location of the hot dust (within 0.2
AU) by combining all interferometric measurements. \citet{defrere11}
further modeled the properties of the hot excess of Vega (both spatial
visibility and SED) and reached several conclusions: the dust
responsible for this hot excess (1) has very steep density and
particle size distributions (density power index less than $-3$ and
particle size power index $\sim -5$), (2) resides from 0.1 AU (dust
sublimation radius with temperature of 1700 K) to less than 0.2 AU
from the star, and (3) has a minimum particle size of 0.01--0.2
\um with a significant fraction of carbonaceous composition (due to
the fact that carbonaceous grains have a much higher sublimation
temperature and lack prominent mineralogical features). Although their
model does include large grains (with maximum size of 1000 \mm) in the
calculation, there is basically no grain with sizes larger than 0.3
\um (the average size particle is 0.27 \mm) with such a steep size
distribution. In other words, the only explanation for such a hot
excess is a population of sub-\um carbonaceous grains located in a
narrow-ring-like region at the dust sublimation radius (0.1 AU).  In
the case of Vega, this hot excess is very different from what we know
about the zodiacal cloud in our solar system, which has a relatively
flat density distribution (density power index of $-$0.34;
\citealt{kelsall98}) and mostly contains large particles with sizes of
$\sim$10--100 \um \citep{fixsen02}. The use of exozodi in this context
as has been referred broadly in the literature for Vega is misleading.

The origin of hot dust population is unknown. One scenario that has
been discussed frequently in the literature is that the hot dust
arises from evaporating comets dynamically perturbed from the outer
cold disks. In the case of Vega, a total mass of $\sim10^{-9}
M_{\earth}$ is required to account for the observed excess emission in
near-IR \citep{defrere11}. The radiation blow-out time scale at 0.1 AU
is on the order of a year, suggesting a very high dust replenishing
rate. This high replenishing rate implies that this phenomenon is
unlikely to be in a static state. If not in steady-state, the hot dust
may be created in transient dynamical events similar to the late heavy
bombardment in our solar system, which was caused by a dynamical
instability of the asteroid belt \citep{strom05}.  However, this kind
of hot excess has also been found around stars that have no detectable
infrared excess indicative of the presence of a planetesimal
population \citep{absil08}, making this scenario unsatisfactory to
explain the hot excess phenomenon.

\citet{kobayashi09} present an analytical model to form a narrow ring
due to sublimation of dust grains drifting radially inward due to P-R
drag. This scenario could work in Vega and Fomalhaut if
the dust drifting inward is refractory; i.e., the asteroid belt discussed in
this paper being the source of the particles. This hypothesis only
requires a P-R dominated disk; no dynamical perturbation is
required. The replenishing rate implies that a total mass of $>$ 0.4
$M_{\earth}$ in the Vega asteroid belt is required to retain such a
rate over the age of the system ($\sim$400 Myr). However, the range of
distances and particle sizes from this drag-in component would result
a much higher excess flux in the mid-IR (see Figure 5 in
\citealt{kobayashi11}). Furthermore, one would expect that the amount of
hot dust should be proportional to the amount of dust in the source
region, i.e., the asteroid belt. The fact that the hot dust in both
systems has a similar fractional luminosity, 5$\times10^{-4}$
\citep{absil06,absil09} while the warm dust in the Fomalhaut system is
$\sim$3 times more than that of the Vega system, argues against this
non-stochastic, transport hypothesis.

There may be an alternative direction for models of the hot dust
component. The basic requirements are (1) a mechanism that prevents
the radiation-pressure blow-out of very small dust grains at very
small astrocentric distances ($\sim$10 stellar radii, i.e.,
the $\sim$1500 K dust that is creating the $K$-band excess); (2) a grain
composition that yields a featureless, roughly Rayleigh-Jeans (or
steeper than Rayleigh Jeans) spectrum between 2 and 10 \um (for
consistency with the SED constraints in Section \ref{stellar}); and
(3) a mechanism for the generation grains of the appropriate
composition.  Although a detailed analysis is beyond the scope of this
paper, these requirements evoke the behavior of nano dust grains
(sizes of ten to a few tens of nm) that are charged by the solar wind
and trapped by the solar magnetic field near the Sun. The underlying
hypothesis is that sungrazing planetesimals deliver silicate-rich
grains to the stellar neighborhood, but when the silicates break down
they yield metal oxides \citep{mann05}. Because of the high abundance
of Mg and Fe in silicates, MgO and FeO are produced profusely in this
process through reactions such as

\begin{displaymath}
\rm MgSiO_3 \rightarrow MgO + SiO + \frac{1}{2}O_2 
\end{displaymath} 

and 

\begin{displaymath}
\rm Fe_2SiO_4 \rightarrow 2FeO + SiO +  \frac{1}{2}O_2 
\end{displaymath}.

\begin{figure}
 \figurenum{10}
 \label{vegahotexcessmodel}
 \plotone{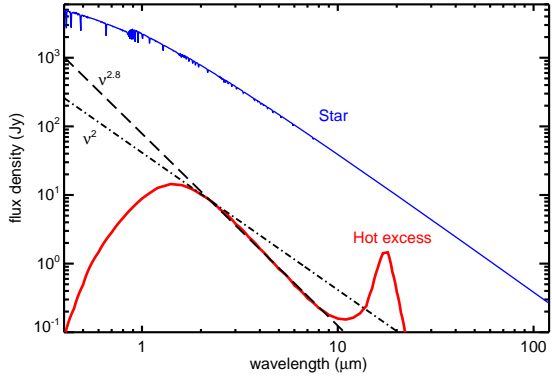}
 \caption{Model SED of the hot excess around Vega using nano oxide
particles. Although the stellar spectrum shown on this plot (blue,
thin, solid line) is the Kurucz atmospheric model (viewed pole-on),
the actual heating spectrum is used the one viewed by the dust along
the equator of Vega. The model spectrum (solid red line) and various
power-law lines (as frequency $\nu$) are normalized to 2.2 \um to account for the 1.29\%
excess at $K$ band.}
\vspace{0.5cm}
\end{figure}

The resulting nano grains are very refractory, for example, MgO has a
melting temperature of 3100 K and boiling temperature of 3873 K,
nearly as high as the sublimation temperature of carbon, 3900 K; FeO
boils at 3687 K. Consequently, MgO, FeO, and composite grains of the
same elements can survive close to the star. We use the hot excess
around Vega as an example to illustrate the resultant SED since the
location for such a hot excess is best constrained among all the $K$\
band excess sources. We adopted the optical constants for a mixed
iron/magnesium oxide, $\rm Mg_{0.6}Fe_{0.4}O$ \citep{henning95} and
computed their absorption and scattering efficiencies using Mie
theory. We computed a model SED (Figure \ref{vegahotexcessmodel}) for
such a narrow ring between 0.18 and 0.2 AU from the star, and composed
of grains with radii of 5--20 nm (0.005--0.02 \mm) in a
size-distribution power index of $-$3.5 using the stellar model
computed by \citet{aufdenberg06} (for dust viewed from the stellar
equator). The dust temperatures for these gains in the region of
0.18--0.2 AU from Vega reach $\sim$2400 K. With the grain density of
4.8 g cm$^{-3}$, a total mass of 2.3$\times$10$^{-9}M_{\earth}$
produces the 1.29\% excess of the star at 2.2 \mm. Toward longer
wavelengths to 10 \mm, the output falls roughly as $\nu^{2.8}$, where
Rayleigh Jeans falls as $\nu^{2}$, that is the spectrum is somewhat
steeper than Rayleigh-Jeans (see Figure
\ref{vegahotexcessmodel}). There is a prominent feature at $\sim$18
\um in our model SED due to the crystalline form of the material
used to determine the optical constants, which might not represent the
actual form of material resulted from sublimation of silicate-rich
planetesimals.  The mass in nano dust is equivalent to about 60
Halley Comet nuclei. Although the conversion of a comet nucleus to
nano dust will not be fully efficient, it should be reasonably high
since the process involves erosion of large grains and so most of the
solid material (which is believed to constitute more than half the
mass of a typical comet nucleus) may be converted into oxides. Thus,
it appears that a plausible number of planetesimals (perhaps no more
than a couple of hundred) would suffice to provide the nano grain
population. Finally, we would like to emphyasize that the model
SED shown in Figure \ref{vegahotexcessmodel} is one example that
satisfies the requirements (as listed previously) in our
hypothesis. Our proposed scenario does not depend on the exact nature
of the material as long as they meet these requirements.

We now consider the magnetic trapping of these dust grains. Vega has a
magnetic field strength of $\sim$1.2 G, with about a quarter in a
dipolar component and the rest in higher order terms
\citep{petit10}. To first order, this is similar to that of
the Sun, with a surface field of about 1 G, dominated by a dipole
but with complex components due to activity. Nano grains in the
vicinity of Vega will acquire electric charge through the
photoelectric effect. This process has been modeled by
\citet{pedersen11}, who find that a 30 nm grain illuminated by an A0
star will reach a level of about 200 $e^-$, or charge per
mass ratio ($Q/m$) $\sim$2 $\times$10$^{-6}$. Smaller grains will reach
higher values of $Q/m$. Assuming a field similar to that of the Sun,
these values are in the range where trapping occurs
\citep{czechowski10}. In summary, the hypothesis that the 2 \um excess
is emitted by nano grains trapped in the magnetic field of Vega does
satisfy the three conditions listed above.  A better understanding of
the behavior of the magnetic field of the star is required to improve
our understanding of whether it is indeed the process that accounts
for this emission. Furthermore, this magnetically trapped, nano dust
model can apply to other stars that show $K$-band hot excesses
through interferometry as long as these stars possess some magnetic
field and can charge these nano particles through photoelectric effect
(for early-type star) or stellar wind (for late-type stars).

\section{Conclusions}

Nearby debris disks, including the debris disk twins: Vega and
Fomalhaut, have been playing an important role in our understanding of
the underlying planetary architectures (planets, and minor bodies)
since their first discovery by {\it IRAS} through infrared
excesses. Much attention has been focused on their large, cold
Kuiper-belt-analog rings because they contain the majority of the
left-over planetesimals and fine debris that covers a large surface
area, making them readily detectable through infrared and
submillimeter observations. The Fomalhaut debris system possesses an
unresolved warm excess first suggested by the {\it Spitzer} resolved
image at 24 \um \citep{stapelfeldt04}. Using the {\it Spitzer} IRS
spectrum centered at this warm excess along with the photometry
measured at the star position from the {\it Herschel} 70 \um and ALMA
850 \um images, we corroborate that the warm ($\sim$170 K) excess
arises from thermal dust emission, and is very unlikely to originate
from stellar wind as suggested by \citet{acke12}. Through resolved
images at multiple wavelengths and mid-infrared spectrum, we
identified for the first time the presence of a warm ($\sim$170 K)
unresolved component in the Vega debris disk, which is clearly
separated from the cold belt. Similar to the one in the Fomalhaut
system, we suggest that this warm component also arises from thermal
dust emission from a planetesimal belt located near the water-frost
line, analogous to the asteroid belt in our solar system. The warm
belts in both systems share many similar characteristics. No extension
at both MIPS 24 and PACS 70 \um (angular radius $\lesssim$3\arcsec)
and the observed dust temperature ($\sim$170 K) place this warm belt
at $\sim$2\arcsec\ from Vega and $\sim$1\farcs5 from Fomalhaut for
blackbody-like grains. Furthermore, the stringent constraints on the
excess levels from 3 to 6 \um around both stars indicate that this
warm belt is not spatially associated with the 2 \um ($K$-band)
excesses inferred from the interferometric observations.

Including the debris disks around $\epsilon$ Eri and HR 8799, the four
nearby debris systems all have spatially separated warm and cold belts
where the observed dust temperatures for the warm belt are at
$\sim$150--170 K, whereas the cold belt temperatures are $\sim$45-50 K
despite the stark differences in stellar luminosity. Similar bi-modal
dust temperature structure is also found in other debris disk
systems. Systems where the location of the cold belts is known through
resolved images either in scattered light or thermal
infrared/submillimeter wavelengths and the association of a separate
warm belt is inferred from spectrophotometric mid-infrared excesses
include the following: HD 10647, HD 15115, HD 107146, and HD 139664
(although not all of them have the warm-belt temperature at $\sim$150
K). Bi-modal temperature distributions inferred from SED analysis for
many other unresolved sources \citep{morales11} argue that the debris
disk structures are, somehow, determined by temperature-sensitive
processes. It is well known that the location of the water-frost line
in protoplanetary disks plays an important role in the formation of
planetesimals. An asteroid belt located near the frost line for these
warm belts seems to imply that they are inherited from the early
evolution of planetary systems.

The $K$-band excesses have been found around roughly a dozen of nearby
main-sequence stars using ground-based interferometric techniques
\citep{absil06,difolco07,absil08,absil09,akeson09}. In some cases,
sources like low-mass companions, ionized stellar wind or stellar
scattered light have been ruled out except for hot ($\sim$1500 K) dust
emission being responsible for the excess emission. Recent $N$-body
simulations done by \citet{bonsor12b} demonstrate the difficulty to
scatter enough small bodies inward by a chain of planets inside a cold
outer belt to sustain the observed level of small grains in Vega and
$\eta$ Crovi. In the case of Vega, the positive detections and
constraints from non-detections using different interferometric
facilities and wavelengths affirm the presence of very small grains in
the very close vicinity ($<$0.2 AU) around the star. The properties of
these hot dust are very different from what we know about the
particles in our own zodiacal cloud; instead, they are more likely to
arise from the phenomenon observed near the Sun where nano particles
trapped in the magnetic field of the star. In our proposed
scenario, nano-size metal oxides originate from the sublimation of
silicate-rich planetesimals, and are charged either via the
photoelectric effect or the stellar wind, and then magnetically
trapped in close proximity to the star. The replenishing rate of
these tiny particles can be very low once they got trapped. Thus, our
scenario does not require a massive reservoir of left-over
planetesimals; therefore, it presumably works for stars that show hot
2 \um excess but without detectable cold dust.

Although the warm belts in the four systems discussed in this paper
are not directly resolved, unlike the cold belts, the orbital ratios
between the outer cold and inner warm belts are roughly $\gtrsim$10
based on the observed dust temperatures. The large gap between the
two planetesimal belts requires a mechanism to maintain it mostly free of
dust. The most plausible explanation is the existence of planetary bodies 
in the gaps, analogous to the HR 8799 and our own solar system. 
From simple chaotic zone calculations and the mass limit on Fomalhaut b, 
we argue that multiple low-mass ($\lesssim$1 $M_J$) planets are required
to maintain such a large gap. A similar argument is also supported for the 
Vega system, suggesting that the widely separated, two-belt debris systems 
are signposts for the presence of multiple low-mass planets. Our
results are in line with the recent result by \citet{wyatt12} where
a positive correlation is identified between the debris disk detection
rate and the presence of exoplanets with mass less than Saturn around
60 nearby G-type stars, and echo the recent RV and {\it
Kepler} results \citep{mayor11,batalha12} that low-mass planets are
more common than the massive ones, and that multiple planet systems are
rather common among exoplanetary systems.

\acknowledgments

We thank the anonymous referee for his/her prompt and constructive
comments. K.Y.L.S. thanks Denis Defr\'ere for his helpful discussion
of the hot excesses. This work is based on observations made with
the {\it Spitzer Space Telescope}, which is operated by the Jet
Propulsion Laboratory, California Institute of
Technology. K.Y.L.S. and K.R.S. are grateful for funding from NASA's
ADAP program (grant No.\ NNX11AF73G).  Support for G.H.R. is provided
by NASA through contract 1255094 and 1256424 issued by JPL/Caltech to
the University of Arizona.  A.M.H. is supported by a fellowship from
the Miller Institute for Basic Research in Science.  A.B. is supported
by the ANR-2010 BLAN-0505-01 (EXOZODI) program. Z.B. is funded by the
Deutsches Zentrum f\"ur Luf- und Raumfahrt (DLR). Partial support for
this work was also provided for Z.B. through Hungarian OTKA Grant No.\
K81966.

\end{document}